\documentclass[conference]{IEEEtran}
\IEEEoverridecommandlockouts

\def\BibTeX{{\rm B\kern-.05em{\sc i\kern-.025em b}\kern-.08em
		T\kern-.1667em\lower.7ex\hbox{E}\kern-.125emX}}

\usepackage{subcaption} 
\usepackage{amsmath,graphicx,amssymb,mathtools,bm}
\usepackage{hyperref}
\usepackage{cite}
\usepackage{amsmath,amssymb,amsfonts}  
\usepackage{textcomp}
\usepackage{xcolor}
\usepackage{verbatim}  
\usepackage{bm}  
\usepackage{mathrsfs} 
 \usepackage{algorithmic} 
\usepackage{booktabs}
\usepackage{textcomp}  
\usepackage{multirow}  
\usepackage{lettrine}   
\usepackage{graphicx}  
\usepackage{color}  
\usepackage{amsmath}
\usepackage{amssymb}
\usepackage{stfloats} 
\usepackage{caption} 
\usepackage{graphicx}
\usepackage{color}  
\usepackage[ruled,linesnumbered]{algorithm2e}

\usepackage{titlesec}
\begin{document}
	\newcommand{\tabincell}[2]{\begin{tabular}{@{}#1@{}}#2\end{tabular}}
   \newtheorem{Property}{\it Property} 
  
 \newtheorem{Proposition}{\bf Proposition}
\newtheorem{remark}{Remark}
\newenvironment{Proof}{{\indent \it Proof:}}{\hfill $\blacksquare$\par}

\title{Movable Subarray-Aided Hybrid Beamforming for Near-Field Multiuser Communications}

\author{Xiangqian Xu\textsuperscript{1},~Songjie Yang\textsuperscript{1},~and Arumugam Nallanathan\textsuperscript{2}, \IEEEmembership{Fellow,~IEEE}\\
\textsuperscript{1}University of Electronic Science and Technology of China, Chengdu, 611731, China\\ \textsuperscript{2}Queen Mary University of London, London E1 4NS, U.K.\\
Email:\{xiangqianxu,yangsongjie\}@std.uestc.edu.cn,~a.nallanathan@qmul.ac.uk

		\thanks{This work has been submitted to the IEEE Globecom for possible publication.}

}

	
	
	
	
\maketitle\thispagestyle{plain}\pagestyle{plain}

\begin{abstract}

Movable antenna (MA)-enabled near-field (NF) communications offer significant potential for 6G, yet existing designs often neglect the practical constraints of hybrid beamforming (HBF) for extremely large-scale MIMO (XL-MIMO). Conversely, MA-aided HBF frequently overlooks the rich NF degrees of freedom (DoFs). This paper proposes a movable subarray (MSA)-aided HBF architecture for NF multiuser systems, which strikes a strategic balance between hardware practicality and spatial flexibility. By coupling MSA movement with HBF, the proposed design simultaneously exploits NF distance-dependent and MSA position-dependent DoFs, enabling highly precise beamfocusing and superior interference mitigation. To alleviate the computational burden, a hybrid planar-spherical wave model is introduced for efficient channel approximation. 
Furthermore, an alternating optimization (AO) algorithm is developed by integrating fractional programming, the alternating direction method of multipliers (ADMM), and projected gradient ascent. Simulation results validate substantial sum-rate gains over fixed-position antenna (FPA) benchmarks.

\end{abstract}
\begin{IEEEkeywords}
Movable antenna (MA), near-field communication, hybrid beamforming, alternating optimization.
\end{IEEEkeywords} 
\section{Introduction}


While extremely large-scale multiple-input multiple-output (XL-MIMO) offers significant multiplexing gains, the associated hardware cost and power consumption present a bottleneck for practical deployment. Movable antenna (MA) technology \cite{zhu2025tutorial,new2024tutorial} has emerged as a promising approach that introduces position-controllable spatial degrees of freedom (DoFs) by enabling local adjustment of antenna elements within continuous spatial regions. This flexibility empowers the transceiver to strategically shape channel realizations, enhancing spectral efficiency by maximizing the utilization of limited radio frequency (RF) chains. Moreover, given the extensive array aperture, signal propagation shifts to the radiative near-field (NF) region characterized by spherical wavefronts. This introduces a new distance dimension to the channel, enriching the spatial DoFs available for transmission. 
By synergistically exploiting distance-dependent DoFs from the near-field and position-dependent DoFs from MAs, MA systems can enable precise 3D beamfocusing and superior interference mitigation in multiuser scenarios\cite{11364019}.

 \begin{figure}[t]
	\begin{center}
		\includegraphics[scale=0.11]{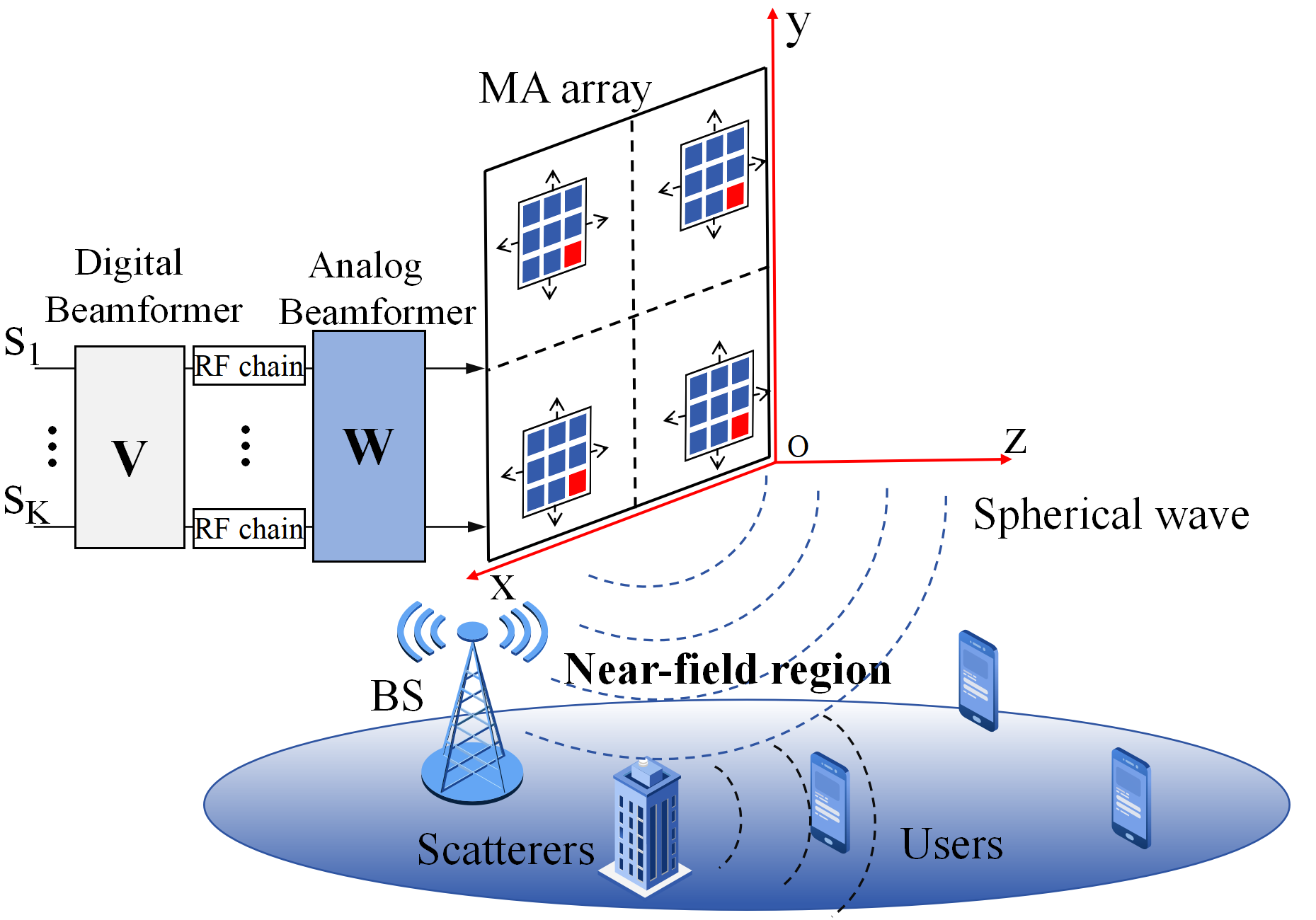}
		\caption{The MSA-aided near-field MU-MISO system. }\label{sys}
	\end{center}
\end{figure}

To fully exploit these capabilities, efforts have been dedicated to both NF and MA communication domains.
In NF communications, research primarily focuses on developing accurate channel models and efficient beamforming architectures to harness distance-dependent DoFs for enhanced spatial multiplexing \cite{liu2023near}. While fully digital (FD) architectures offer optimal performance, their prohibitive hardware and power consumption in XL-MIMO systems necessitate hybrid beamforming (HBF), making the design of NF-HBF schemes a key research priority \cite{10045774}. Concurrently, MA systems have gained traction for their ability to proactively reshape wireless channels, leading to considerable interest in both communication and sensing applications\cite{jiang2025movable,lyu2025movable}. However, the practical implementation of element-level MAs in XL-MIMO is hindered by high energy consumption and mechanical movement delays, thereby motivating the adoption of movable subarrays (MSAs) that naturally align with HBF architectures \cite{ning2025movable}. For instance, \cite{zhang2025movable} investigated MSA-aided multiuser HBF, yet this work was restricted to far-field assumptions. Preliminary studies have further revealed the benefits of the combination of MA and NF in multiuser communications, including minimizing transmit power \cite{ding2024near}, maximizing the minimum signal-to-interference-plus-noise ratio (SINR)\cite{zhu2025movable}, and maximizing the weighted sum-rate\cite{pi2025movable}, compared to their fixed-position antenna (FPA) counterparts. Nonetheless, existing MA-enabled NF multiuser designs do not incorporate HBF structures, which restricts their practicality in XL-MIMO systems.

 To bridge this gap, this paper proposes an MSA-aided HBF architecture for NF multiuser multiple-input single-output (MU-MISO) systems. Unlike prior works, we formulate a joint optimization problem for the digital beamformer, analog beamformer, and MSA positions to maximize the sum-rate. To tackle the significant non-convexity and variable coupling, we develop an efficient alternating optimization (AO) algorithm integrating fractional programming (FP), the alternating direction method of multipliers (ADMM), and projected gradient ascent. Simulation results demonstrate that the proposed scheme effectively harnesses the synergy of NF and MSA position-dependent DoFs, significantly outperforming conventional FPA systems.



\section{System Model and Problem Formulation}\label{sys_model}
 \subsection{Signal Model}
  
We consider a  MSA-aided near-field downlink  MU-MISO system, as illustrated in Fig. \ref{sys}.  The BS is equipped with a large 2D MA array composed of $M$ MSAs, and user $k$ $(k \in \mathcal{K} \triangleq\{1,\ldots, K\})$ is equipped with a FPA, while each MSA is restricted to moving within a designated fixed area. Without any loss of generality, we assume that each MSA is
a square uniform planar array (UPA) consisting of $N$ antennas. 
The BS adopts a sub-connected HBF
architecture, which is  equipped with $N_{RF}$ ($N_{RF}=M$) RF chains. In each UPA, antennas are individually connected to PSs and a dedicated RF chain, forming a collective MSA unit. 


For clarity, a 3D Cartesian coordinate system is established with its origin at the BS. The $x$ and $y$ axes align with the horizontal and vertical dimensions of the 2D MA array, respectively, while the $z$-axis is perpendicular to the array plane. We define the movable region for each MSA as $\mathcal{C}_m, m=1,\ldots,M$, within which the MSA can move freely. For each MSA, the first antenna is designated as the reference point and is highlighted in red. The coordinates of $M$ MSA reference point can be described by $\mathbf{t}\triangleq\left[\mathbf{t}_1^T,\ldots, \mathbf{t}_m^T,\ldots, \mathbf{t}_M^T\right]^T \in \mathbb{R}^{3 M \times 1}$, where $\mathbf{t}_m=\left[x_m, y_m, 0\right]^T $. The $(n, m)$-th
antenna element of the 2D MA array is located at $\mathbf{t}_{n,m}=\left[x_{n,m}, y_{n,m}, 0\right]^T$, representing the $n$-th antenna of the $m$-th MS, and $\tilde{\mathbf{t}} \triangleq \left[\mathbf{t}_{1,1}^T,\ldots, \mathbf{t}_{n,m}^T,\ldots, \mathbf{t}_{N,M}^T\right]^T$. 
Furthermore, each user is assumed to receive signals through one line-of-sight (LoS) path from the BS and $(N_p-1)$ non-line-of-sight (NLoS) paths, due to multipath propagation resulting from environmental scattering. 
 The location of the $k$-th user in 3D coordinate is represented by $\mathbf{r}_{k, 1}$, while the  3D location of the $l$-th scatterer affecting the $k$-th user's channel is denoted by $\mathbf{r}_{k, l}$, $2 \leq l \leq N_p$. Then, the NF multipath channel \cite{liu2023near} of $k$-th user could be modeled by
\begin{equation}\label{ste}
	\mathbf{h}_k(\tilde{\mathbf{t}})=\sum_{i=1}^{N_p} \beta_{k, i} \tilde{\mathbf{b}}_{k,i}\left(\tilde{\mathbf{t}}, \mathbf{r}_{k, i}\right),
\end{equation}
where $\beta_{k, i}$ is the $i$-th complex path gain of the $k$-th UE. $\tilde{\mathbf{b}}_{k,i}$ $\in \mathbb{C}^{MN \times 1}$ denotes the near-field steering vector, with each element given by $\tilde{\mathbf{b}}_{k,i}[nm]$ = $\mathrm{e}^{\mathrm{-j} \frac{2 \pi}{\lambda}\left\|\mathbf{t}_{n,m}-\mathbf{r}_{k, i}\right\|_2}$. 

To reduce the complexity associated with subsequent position optimization designs, inspired by \cite{8356240}, we approximate the spherical wave channel using a hybrid planar-spherical wave model. This model incorporates spherical-wave propagation between subarrays while assuming planar-wave propagation within each subarray. Thus,
the NF steering vector $\tilde{\mathbf{b}}_{k,i}\left(\tilde{\mathbf{t}}, \mathbf{r}_{k, i}\right)$ in (\ref{ste}) can be constructed by 
\begin{equation}
	\tilde{\mathbf{b}}_{k,i}\left(\tilde{\mathbf{t}}, \mathbf{r}_{k, i}\right)\triangleq \mathbf{b}_{k,i}(\mathbf{t},\mathbf{r}_{k,i}) \otimes \mathbf{a}_{k,i},
\end{equation}
where $\mathbf{b}_{k,i}$ $\in \mathbb{C}^{M \times 1}$ denotes the phase vector among subarrays over the $i$-th multipath for $k$-th user, satisfying $\mathbf{b}_{k,i}[m]=\mathrm{e}^{\mathrm{-j} \frac{2 \pi}{\lambda}\left\|\mathbf{t}_{m}-\mathbf{r}_{k, i}\right\|_2}$. In addition,
the planar-wave
assumption is appropriate within each subarray, and $\mathbf{a}_{k,i}$  is the array response vector of the far-field UPA\cite{10045774}.

Consider a narrow-band system, the received signal of the $k$-th user can be modeled as 
\begin{equation}
	y_k=\mathbf{h}^H_k(\mathbf{t}) \mathbf{W} \sum_{j=1}^K \mathbf{v}_j s_j+n_k, k=1,2, \ldots, K,
\end{equation}
where $s_j$ $\in \mathbb{C}$ is the transmit symbol for the $j$-th user, which is independent both within a user stream and across different users, with $\mathbb{E}\left\{\left|s_j\right|^2\right\}=1$ and $n_k$ $\sim \mathcal{C N}\left(0, \sigma_k^2\right)$ is additive Gaussian white noise. $\mathbf{V} \triangleq [\mathbf{v}_1,\ldots,\mathbf{v}_k,\ldots,\mathbf{v}_K] \in \mathbb{C}^{M \times K}$ denotes the digital beamformer, where  $\mathbf{v}_k \in \mathbb{C}^{M\times 1}$ is the digital beamformer vector for the $k$-th user. Due to the sub-connected architecture, the analog beamformer $\mathbf{W} \in \mathbb{C}^{MN \times M}$ takes the form $	\mathbf{W}\triangleq\operatorname{diag}\left(\mathbf{w}_{1},\ldots, \mathbf{w}_{m}, \ldots, \mathbf{w}_{M}\right)$, where each element of $\mathbf{w}_{m}\in \mathbb{C}^{N \times 1}$ has unit amplitude, i.e. $\left|\mathbf{w}_m(i)\right|=1, i=1,\ldots,N$. Accordingly, the sum-rate of the considered system is given by
 \begin{equation}\label{sum}
	R=\sum_{k=1}^K \log _2\left(1+\frac{\left|\mathbf{h}^H_k(\mathbf{t}) \mathbf{W} \mathbf{v}_{ k}\right|^2}{\sum_{j \neq k}\left|\mathbf{h}^H_k(\mathbf{t}) \mathbf{W} \mathbf{v}_{ j}\right|^2+\sigma_k^2}\right).
\end{equation}

 
  \subsection{Problem Formulation}
  To maximize the sum-rate of the considered system, the optimization
  problem can be formulated as
 \begin{equation}\label{P00}
 	\begin{aligned}
 	(\text{P1} ) :	\max _{\mathbf{W}, \mathbf{V}, \mathbf{t}} &  R \\
 		\text { s.t. } &\mathrm{C}_1:  \mathbf{t}_{m} \in \mathcal{C}_m, \forall m  \\
 		&\mathrm{C}_2: 	\left|\mathbf{w}_m(i)\right|=1, \forall m,  i \\
 		&\mathrm{C}_3: \left\|\mathbf{W} \mathbf{V}\right\|_F^2 \leq P,
 	\end{aligned}
 \end{equation}
 where $P$ is the total transmit power and $ \mathcal{C}_m\triangleq \{ (x_m, y_m) \in \mathbb{R}^2 \mid a \leq x_m \leq b, \; c \leq y_m \leq d \}$, with $a$ and $b$ indicating the boundary limits in the $x$-direction, and $c$ and $d$ in the $y$-direction. Accordingly, constraint $\mathrm{C}_1$ imposes limits on the allowable movement regions for each subarray. Constraints $\mathrm{C}_2$ and $\mathrm{C}_3$ enforce unit-modulus constraint and transmit power constraint, respectively. The inherent non-convexity of (P1) and the tight coupling among the variables $\mathbf{W}$, $\mathbf{V}$, and $\mathbf{t}$ make its direct solution challenging. To tackle this issue, we decompose (P1) into three subproblems. Within an AO framework, we leverage FP, the ADMM, and projected gradient ascent to iteratively optimize the digital beamformer, the analog  beamformer, and the MSAs position. The details of the proposed AO algorithm are elaborated in Section \ref{S3}.

 
 
 

\section{Proposed Algorithm}\label{S3}

  \subsection{Transformation of Objective Function                }
Inspired by the use of FP in \cite{fp1} to solve multiple-ratio problems, we first apply the Lagrangian dual transform to separate the ratio terms from the logarithm in (\ref{sum}). Then, the objective function of ($\text{P1}$) can be reformulated as
 \begin{equation}\label{f1}
f_1\left(\mathbf{W}, \mathbf{V},\mathbf{t}, \boldsymbol{\eta}\right)= \sum_{k=1}^K \log _2\left(1+\eta_k\right)- f_2(\mathbf{W}, \mathbf{V}, \mathbf{t}, \boldsymbol{\eta}, \boldsymbol{\mu}), 
\end{equation}
 where $\boldsymbol{\eta}\triangleq\left[\eta_1,  \ldots,\eta_k, \ldots, \eta_K\right]$ is the auxiliary variable vector, which in turn defines another auxiliary vector $\boldsymbol{\mu}\triangleq\left[\mu_1, \ldots, \mu_k, \ldots, \mu_K\right]$. The term $f_2(\mathbf{W}, \mathbf{V}, \mathbf{t}, \boldsymbol{\eta}, \boldsymbol{\mu})$ in the above equation can be further expressed as:
 
  \begin{equation}\label{F2}
  		 -\sum_{k=1}^K\left|\mu_k\right|^2 \mathbf{A}_k+\sum_{k=1}^K 2 \sqrt{1+\eta_k} \Re\left\{\mu_k^* \mathbf{h}_k^H(\mathbf{t}) \mathbf{W} \mathbf{v}_k\right\},
  \end{equation}
 when the auxiliary variables $\eta_k$ and $\mu_k$  both attain the following optimal values:
\begin{equation}\label{p1}
	\eta_k^{\star}=\frac{\left|\mathbf{h}^H_k(\mathbf{t}) \mathbf{W} \mathbf{v}_{ k}\right|^2}{\sum_{j \neq k}\left|\mathbf{h}^H_k(\mathbf{t}) \mathbf{W} \mathbf{v}_{ j}\right|^2+\sigma_k^2}, \forall k ,
\end{equation}
and
\begin{equation}\label{p2}
	\mu_k^{\star}=\frac{\sqrt{1+\eta_k} \mathbf{h}^H_k(\mathbf{t}) \mathbf{W} \mathbf{v}_{ k}}{A_k},\forall k ,
\end{equation}
where $A_k=\sum_{j=1}^K\left|\mathbf{h}^H_k(\mathbf{t}) \mathbf{W} \mathbf{v}_j\right|^2+\sigma_k^2$ .

 
   \subsection{Optimization of the Digital Beamformer     }
 
 In this subsection, we detail the steps for obtaining $\mathbf{V}$ given the parameters $\left\{\mathbf{W},\mathbf{t},\boldsymbol{\eta},\boldsymbol{\mu}\right\}$. Dropping the constant
 terms of the function $f_1$, the problem of designing the optimal digital beamformer can be formulated as
    \begin{equation}
  	\begin{aligned}
  		(\text{P2} ) :	\max _{\mathbf{W}, \mathbf{V}, \mathbf{t}} &  f_2(\mathbf{W}, \mathbf{V}, \mathbf{t}, \boldsymbol{\eta},\boldsymbol{\mu}) \\
  		\text { s.t. } 
  		&\mathrm{C}_3: \left\|\mathbf{W} \mathbf{V}\right\|_F^2 \leq P.
  	\end{aligned}
  \end{equation}
  
  In light of the preceding analysis, the effective channel for the $k$-th UE can be expressed as $\bar{\mathbf{h}}_k(\mathbf{t})\triangleq\mathbf{W}^H \mathbf{h}_k(\mathbf{t}) $. Furthermore, we define $\mathbf{B}_k=\left|\mu_k\right|^2 \bar{\mathbf{h}}_k(\mathbf{t}) \bar{\mathbf{h}}^H_k(\mathbf{t}),\forall k$. Then
  the optimal solution to $(\text{P2} )$ can be obtained through the
  method of Lagrange multipliers. By introducing a multiplier
  $\lambda$ associated with the power constraint $\mathrm{C}_3$, we can construct
  the Lagrangian function as follows:
\begin{equation}
	L_\lambda=f_2(\mathbf{W}, \mathbf{V}, \mathbf{t}, \boldsymbol{\eta}, \boldsymbol{\mu})+\lambda\left(P-\sum_{h=1}^K\left\|\mathbf{W} \mathbf{v}_k\right\|_2^2\right) .
\end{equation}

The optimal solution for $\mathbf{v}_k$ can be found by setting the partial derivatives of $L_\lambda$ with respect to $\mathbf{v}_k$ and $\lambda$ equal to zero, i.e., $\frac{L_\lambda}{\partial \lambda}=0$ and $\frac{L_\lambda}{\partial \mathbf{v}_k}=0, \forall k$. This consequently defines the optimal digital beamformer of the $k$-th user as
 \begin{equation}\label{147}
	\mathbf{v}_k^{\star}=\left(\sum_{j=1}^K\mathbf{B}_j+ \lambda^{\star} \mathbf{W}^H \mathbf{W}\right)^{\dagger} \sqrt{\left(1+\eta_k\right)} \bar{\mathbf{h}}_k(\mathbf{t}) \mu_k, \quad \forall k,
\end{equation}
where the optimal multiplier $\lambda^{\star}$ is introduced to satisfy the power constraint and can be efficiently determined using bisection search. In (\ref{147}), the parameter $\lambda$ is defined as $\lambda=\min \left\{\lambda \geq 0: \left\|\mathbf{W} \mathbf{V}\right\|_F^2 \leq P\right\}.$

 \subsection{Optimization of the Analog Beamformer       }
 In this subsection, we search for optimal $\mathbf{W}^{\star}$ with the fixed variables $\left\{\mathbf{V},\mathbf{t}\right\}$. To begin with,  we employ fundamental mathematical transformations to derive the equation $\mathbf{h}^H_k(\mathbf{t}) \mathbf{W} \mathbf{v}_{ j} = \mathbf{h}^H_k(\mathbf{t}) \operatorname{diag}\left(\mathbf{\Phi} \mathbf{v}_j\right) \hat{\mathbf{w}}, \forall k, j$, where $\mathbf{\Phi}=\text{diag}(\mathbf{1}_1, \dots,\mathbf{1}_m, \dots, \mathbf{1}_{M})\in\mathbb{C}^{MN \times M}$ and $\hat{\mathbf{w}}\triangleq\left[\mathbf{w}_1^T, \mathbf{w}_2^T, \ldots, \mathbf{w}_{M}^T\right]^T\in\mathbb{C}^{MN\times1}$ respectively, and $\mathbf{1}_m\in\mathbb{C}^{N \times 1}$ is a vector with all elements equal to 1. Further, we can define $\hat{\mathbf{h}}_{k,j}(\mathbf{t}) \triangleq\operatorname{diag}\left(\mathbf{\Phi} \mathbf{v}_j\right)^H \mathbf{h}_k(\mathbf{t}) \in \mathbb{C}^{MN \times 1}, \forall k, j$.
Given that $\boldsymbol{\eta}$ and $\boldsymbol{\mu}$ are derived from equations (\ref{p1}) and (\ref{p2}), the constant term in equation (\ref{f1}) can be omitted, which allows us to reformulate the objective function $f_1$ as follows:
 \begin{equation}
 	\begin{aligned}
 		f_3(\hat{\mathbf{w}})= & \sum_{k=1}^K 2 \sqrt{1+\eta_k} \Re\left\{\mu_k^* \hat{\mathbf{h}}_{k, k}^H(\mathbf{t}) \hat{\mathbf{w}}\right\} \\
 		& -\sum_{k=1}^K\left|\mu_k\right|^2\left(\sum_{j=1}^K\left|\hat{\mathbf{h}}_{k, j}^H(\mathbf{t}) \hat{\mathbf{w}}\right|^2+\sigma_k^2\right) .
 	\end{aligned}
 \end{equation}
 
 
 Furthermore, to address the constant modulus constraint in (\ref{P00}) while facilitating subsequent optimization of the MA positions without being influenced by the beamforming design. The constraint $\mathrm{C}_2$ of $(\text{P1} )$ is replaced by the less restrictive peak power constraint $\left|\hat{\mathbf{w}}(i)\right|\leq1, i=1,2,\ldots,MN$.
  Eventually, dropping the constant terms and applying the equality $\left|\hat{\mathbf{h}}^H_{k,j}(\mathbf{t}) \hat{\mathbf{w}}\right|^2$ =$\hat{\mathbf{w}}^H \hat{\mathbf{h}}_{k,j}(\tilde{\mathbf{t}}) \hat{\mathbf{h}}^H_{k,j}(\tilde{\mathbf{t}}) \hat{\mathbf{w}}$, $(\text{P1} )$ is equivalent to $(\text{P3} )$ for fixed $\boldsymbol{\eta}$ and  $\boldsymbol{\mu}$:
 \begin{equation}\label{P2}
	\begin{aligned}
		(\text{P3} ): 	\max _{\hat{\mathbf{w}}} & f_4(\hat{\mathbf{w}})= 2 \Re\left\{ \mathbf{r}^H\hat{\mathbf{w}}-\hat{\mathbf{w}}^H \mathbf{T} \hat{\mathbf{w}}\right\} \\
		\text { s.t. } &\mathrm{C}_4: \sum_{k=1}^K \|\hat{\mathbf{v}}_k \hat{\mathbf{w}}\|_2^2 \leq P \hspace{0.5cm} \mathrm{C}_5:  \left|\hat{\mathbf{w}}(i)\right|\leq1, \forall i,\\
	\end{aligned}
\end{equation}
 where $\mathbf{T}\triangleq\sum_{k=1}^K\left|\mu_k\right|^2\left(\sum_{j=1}^K \hat{\mathbf{h}}_{k,j}(\mathbf{t}) \hat{\mathbf{h}}^H_{k,j}(\mathbf{t})\right)$, $\hat{\mathbf{v}}_k\triangleq\operatorname{diag}\left(\mathbf{\Phi} \mathbf{v}_k\right)$ and $\mathbf{r} \triangleq\sum_{k=1}^K\left(\sqrt{1+\eta_k} \mu_k\hat{\mathbf{h}}_{k,k}(\mathbf{t})\right) $. Specifically, modifications in $\mathrm{C}_5$ exclusively influence $\mathbf{W}$, while the solutions for $\mathbf{V}$ and $\mathbf{t}$ remain unchanged. With variables $\left\{\mathbf{V},\mathbf{t}\right\}$ fixed, we first introduce two sets of auxiliary variables, which are $\boldsymbol{\kappa}_i=\hat{\mathbf{w}}(i)$ and $\boldsymbol{\zeta}_k = \hat{\mathbf{v}}_k \hat{\mathbf{w}}$. To ease the notation, we define  $\boldsymbol{\kappa}=\left\{\boldsymbol{\kappa}_1,\boldsymbol{\kappa}_2,\ldots,\boldsymbol{\kappa}_{MN}\right\}$ and $\boldsymbol{Z}=\left\{\boldsymbol{\zeta}_1,\boldsymbol{\zeta}_2,\ldots,\boldsymbol{\zeta}_{K}\right\}$.
 Then, the problem $(\text{P3} )$ can be equivalently expressed as
 \begin{equation}
 	\begin{aligned}
 		(\text{P4} ): \min _{\hat{\mathbf{w}},\boldsymbol{\kappa}, \mathbf{Z}} & -f_4(\hat{\mathbf{w}})+\mathbb{I}_{\mathcal{C}}(\boldsymbol{\kappa})+\mathbb{I}_{\mathcal{D}}(\boldsymbol{Z}) \\
 		\text { s.t. } & \mathrm{C}_6:\boldsymbol{\kappa}-\hat{\mathbf{w}}=0 \hspace{0.5cm}\mathrm{C}_7:\boldsymbol{\zeta}_k-\hat{\mathbf{v}}_k \hat{\mathbf{w}}=0, \forall k.
 	\end{aligned}
 \end{equation}

 We define the feasible region of constraint $\left|\boldsymbol{\kappa}_i\right|\leq1$ , $\forall i$ and $\sum_{k=1}^K \|\boldsymbol{\zeta}_k\|_2^2 \leq P$, as $\mathcal{C}$ and $\mathcal{D}$ respectively, and 
 their indicative functions\cite{10161727} are
 \begin{equation}
	\mathbb{I}_\mathcal{C}(\boldsymbol{\kappa})= \begin{cases}0, & \text { if } \boldsymbol{\kappa} \in \mathcal{C} \\ +\infty, & \text { otherwise }\end{cases}
 \end{equation}
 
 and
 
 \begin{equation}
 	\mathbb{I}_\mathcal{D}(\boldsymbol{\zeta})= \begin{cases}0, & \text { if } \boldsymbol{\zeta} \in \mathcal{D} \\ +\infty, & \text { otherwise }\end{cases}
 \end{equation}
Then, we have the augmented Lagrangian of $(\mathbf{P} 4)$:
 \begin{equation}
 	\begin{aligned}
 		\mathcal{L}(\hat{\mathbf{w}},\boldsymbol{\kappa}, \mathbf{Z}, \boldsymbol{x}, \boldsymbol{z})= & -f_4(\hat{\mathbf{w}})+\mathbb{I}_{\mathcal{C}}(\boldsymbol{\kappa})+\mathbb{I}_{\mathcal{D}}(\boldsymbol{Z}) \\
 		& +\frac{\rho}{2}\|\boldsymbol{\kappa}-\hat{\mathbf{w}}+\boldsymbol{x}\|_2^2 \\
 		&+\frac{\rho}{2} \sum_{k=1}^K\left\|\boldsymbol{\zeta}_k-\hat{\mathbf{v}}_k \hat{\mathbf{w}}+\boldsymbol{z}_k\right\|_2^2,
 	\end{aligned}
 \end{equation}
 where  $\rho> 0$  is the penalty parameter, $\boldsymbol{x}$ and $\boldsymbol{z}\triangleq \left\{\boldsymbol{z}_1, \boldsymbol{z}_2, \ldots, \boldsymbol{z}_K\right\}$ are the scaled dual variables for constraints $\mathrm{C}_6$ and $\mathrm{C}_7$, respectively. Accordingly, we can directly apply the ADMM method \cite{10161727} to derive the optimal closed-form solution for $\hat{\mathbf{w}}$ as follows:

 \begin{equation}\label{p7}
 	\hat{\mathbf{w}}^{\star}=\left(2 \mathbf{T}+\rho \mathbf{I}_{MN}+\rho \hat{\mathbf{V}}\right)^{-1} \mathbf{\Theta},
 \end{equation}
 where  $\mathbf{\Theta}=\left(2 \mathbf{r}+\rho\left(\boldsymbol{\kappa}+\boldsymbol{x}\right)+\rho \sum_{k=1}^K \hat{\mathbf{v}}^H_k\left(\boldsymbol{\zeta}_k+\boldsymbol{z}_k\right)\right)$ and $\hat{\mathbf{V}} = \sum_{k=1}^K \hat{\mathbf{v}}^H_k \hat{\mathbf{v}}_k$.
 Notably,
 while $(\text{P3} )$ can be directly tackled using CVX, the computational complexity arising from the extensive antenna deployment in XL-MIMO systems is unacceptable.
  Based on equation (\ref{p7}), the optimal analog beamforming matrix  $\mathbf{W}$ is given by
      \begin{equation}\label{p71}
    \mathbf{W}^{\star}=\operatorname{diag}(\hat{\mathbf{w}}) \mathbf{\Phi},
    \end{equation}

  \subsection{Optimization of MSAs position      }
 In this subproblem, we aim to optimize $\mathbf{t}_m$ in $(\text{P1} )$ with given $\left\{\mathbf{V},\mathbf{W},\left\{\mathbf{t}_i, i \neq m\right\}_{i=1}^M\right\}$, $\forall m \in\mathcal{M}=\{1,2, \ldots, M\}$. For notation simplicity, we define $\boldsymbol{\omega}_k\in \mathbb{C}^{ MN \times 1}$ as follows:
  \begin{equation}
\boldsymbol{\omega}_k=\mathbf{W} \mathbf{v}_k\triangleq\left[\boldsymbol{\omega}_{k,1}^T, \ldots,\boldsymbol{\omega}_{k,m}^T, \ldots, \boldsymbol{\omega}_{k,M}^T\right]^T,
 \end{equation}
where $\boldsymbol{\omega}_{k,m} \in \mathbb{C}^{N \times 1}$ consists of the elements from the $((m-1)N + 1)$-th to the $(mN)$-th element of $\boldsymbol{\omega}_k$. Given $\left\{\mathbf{V},\mathbf{W},\boldsymbol{\eta},\boldsymbol{\mu}\right\}$, we can reformulate the objective function $f_2$ in (\ref{F2}) as
\begin{equation}
		\begin{aligned}
		f_5( \mathbf{t})= & \sum_{k=1}^K 2 \sqrt{1+\eta_k} \Re\left\{\mu_k^* \mathbf{h}_k^H(\mathbf{t}) \boldsymbol{\omega}_k\right\} -\hat{f_5}( \mathbf{t})\\
	\end{aligned}
\end{equation}
where $\mathbf{h}_k(\mathbf{t})\triangleq\left[\mathbf{c}_k^H(\mathbf{t}_1), \ldots,\mathbf{c}_k^H(\mathbf{t}_m), \ldots, \mathbf{c}_k^H(\mathbf{t}_M)\right]^H$ with $\mathbf{c}_k(\mathbf{t}_m)=\sum_{i=1}^{N_p} \beta_{k, i} \mathbf{b}_{ki}[m] \mathbf{a}_{k,i}\left(\theta_{k, i}, \phi_{k, i}\right)\in \mathbb{C}^{ N \times 1}$ and $\hat{f_5}( \mathbf{t})=\sum_{k=1}^K\left|\mu_k\right|^2 ((\mathbf{h}^H_k(\mathbf{t})\sum_{j=1}^K\boldsymbol{\omega}_j\boldsymbol{\omega}_j^H\mathbf{h}_k(\mathbf{t}))^*+\sigma_k^2)$. Further, we define $\sum_{j=1}^K\boldsymbol{\omega}_j\boldsymbol{\omega}_j^H$ as follows:
\begin{equation}
	\sum_{j=1}^K\boldsymbol{\omega}_j\boldsymbol{\omega}_j^H\triangleq\left[\begin{array}{cccc}
		\mathbf{X}_{11} &   \ldots & \mathbf{X}_{1M} \\
		
		\vdots &\ddots& \vdots \\
		\mathbf{X}_{M1} & \ldots & \mathbf{X}_{MM}
	\end{array}\right]_{MN \times MN},
\end{equation}
where $\mathbf{X}_{ij}=\sum_{k=1}^K\boldsymbol{\omega}_{k,i}\boldsymbol{\omega}_{k,j}^H\in \mathbb{C}^{ N \times N},i,j=1,\ldots,M$. It is noteworthy that $\mathbf{X}_{ij}$ and $\mathbf{X}_{ji}$ are conjugate transposes, while $\mathbf{X}_{ii}$ is a positive semi-definite matrix. Thus, we can redefine the objective function $f_5$ as follows:
\begin{equation}\label{f51}
	\begin{aligned}
		f_5^*( \mathbf{t})= & \sum_{k=1}^K 2 \sqrt{1+\eta_k} \Re\left\{\mu_k^* \sum_{m=1}^M\mathbf{c}_k^H(\mathbf{t}_m)\boldsymbol{\omega}_{k,m}\right\} \\
		& -\sum_{k=1}^K\left|\mu_k\right|^2 (\sum_{i=1}^M(\mathbf{c}_k^H(\mathbf{t}_i)\sum_{m=1}^M\mathbf{X}_{im}\mathbf{c}_k(\mathbf{t}_m))^*+\sigma_k^2).
	\end{aligned}
\end{equation}

Observe that (\ref{f51}) clearly decouples the position variables for all $M$ MSAs, i.e., $\left\{\mathbf{t}_m\right\}_{m=1}^M$, which streamlines the following optimization process of $\mathbf{t}_m$. Consequently, the optimization problem of solving $\mathbf{t}_m^{\star}$ can be expressed as

\begin{equation}\label{P4}
	\begin{aligned}
		(\text{P5} ): 	\max _{\mathbf{t}_m} & f_6(\mathbf{t}_m)= \sum_{k=1}^K2  \Re\left\{ \mathbf{c}_k^H(\mathbf{t}_m)\mathbf{f}_k\right\} \\
		& -\sum_{k=1}^K\left|\mu_k\right|^2\mathbf{c}_k^H(\mathbf{t}_m)\mathbf{X}_{mm}\mathbf{c}_k(\mathbf{t}_m) \\		
		\text { s.t. } &\mathrm{C}_1:  \mathbf{t}_{m} \in \mathcal{C}_m, \forall m  ,
	\end{aligned}
\end{equation}
where $\mathbf{f}_k=\sqrt{1+\eta_k}\mu_k^*\boldsymbol{\omega}_{k,m}-\left|\mu_k\right|^2\sum_{i\neq m}\mathbf{X}_{mi}\mathbf{c}_k(\mathbf{t}_i)$.
Given that constraint $\mathrm{C}_1$ is a simple linear constraint, we can employ the gradient ascent method  to address the above optimization problem $(\text{P5} )$. 
Based on (\ref{P4}), the gradient of the function $f_6(\mathbf{t}_m)$ is represented as follows:
\begin{equation}\label{p8}
	\begin{aligned}
		\nabla_{\mathbf{t}_{m}} f_6(\mathbf{t}_m)=2 \sum_{k=1}^K \mathcal{R}\left\{\frac{\partial \mathbf{c}^H_k\left(\mathbf{t}_{m}\right)}{\partial \mathbf{t}_{m}}\mathbf{f}_k \right. \\
		\left.\quad-\left|\mu_k\right|^2\frac{\partial \mathbf{c}^H_k\left(\mathbf{t}_{m}\right)}{\partial \mathbf{t}_{m}}\mathbf{X}_{mm}\mathbf{c}_k(\mathbf{t}_m)\right\},
	\end{aligned}
\end{equation}
where  $\frac{\partial \mathbf{c}_k\left(\mathbf{t}_{m}\right)}{\partial \mathbf{t}_{m}}=-j \frac{2 \pi}{\lambda}  \sum_{i=1}^{N_p}\beta_{k, i}
\mathbf{b}_{ki}[m]\mathbf{a}_{k,i}\frac{(\boldsymbol{t}_m-\boldsymbol{r}_{k,i})}{\left\|\boldsymbol{t}_m-\boldsymbol{r}_{k,i}\right\|_2}$.
 After each iteration, if the updated parameter $\mathbf{t}_{m}^{(n+1)}$ of the $n$-th iteration lies outside the feasible region $\mathcal{C}_m$, it is corrected to fall within this region using a projection function. This projection can be defined as: $\mathbf{t}_m^{(n+1)} = P_{\mathcal{C}_m}(\mathbf{t}_m^{(n+1)})$, where
 the projection function $P_{\mathcal{C}_m}$ can be expressed as
 
 
 \begin{equation}
P_{\mathcal{C}_m}\left(\mathbf{t}_{m}^{(n+1)}\right) = \begin{bmatrix}
	\max(a, \min(x_m^{(n+1)}, b)) \\
	\max(c, \min(y_m^{(n+1)}, d))
	\end{bmatrix}.
\end{equation}

   \begin{algorithm} [ht]
 	\caption{Proposed algorithm for problem (\ref{P00}) } 
 	\label{alg1} 
 	\begin{algorithmic}[1] 
 		\REQUIRE	
 		Obtain CSI feedback from K users.
 		\ENSURE 
 		$\mathbf{W}$, $\mathbf{V}$ and $\mathbf{t}$.
 		
 		\STATE Initialization: $n=0$ , $\mathbf{W}^0$, $\mathbf{V}^0$, $\mathbf{t}^0$,  $\tau^0$, $\epsilon_1$, $\epsilon_2$, $\epsilon_3$,$I$.
 		\WHILE{$\left|\mathrm{R}^{n+1}-\mathrm{R}^{n}\right|^2 \geq \epsilon_1$, and $n<I$}
 		\STATE Fix $\mathbf{W}^n$, $\mathbf{V}^n$, $\mathbf{t}^n$.
 		\STATE Update $\boldsymbol{\eta}, \boldsymbol{\mu}$ sequentially using (\ref{p1}), (\ref{p2}).
 		\STATE Obtain $\mathbf{V}^{n+1}$ using (\ref{147}).
 		\STATE Fix $\mathbf{W}^n$, $\mathbf{V}^{n+1}$, $\mathbf{t}^n$.
 		\STATE Obtain $\mathbf{W}^{n+1}$ using (\ref{p71}).
 		\STATE Fix $\mathbf{W}^{n+1}$, $\mathbf{V}^{n+1}$, $\mathbf{t}^n$.
 		\FOR{$m=1 \rightarrow M$}
 		\STATE Fix $\left\{\mathbf{t}_i, i \neq m\right\}_{i=1}^M$, obtain $\nabla_{\mathbf{t}_{m}} f_6(\mathbf{t}_m^{(n)})$ via (\ref{p8}).
 		\STATE Initialize the step size  $\tau=\tau^0$.
 		\WHILE{$\mathbf{t}_{\text{temp}}^{(n)} \notin\mathcal{C}_m \text { or } f_6(\mathbf{t}_{\text{temp}}^{(n)}) \leq f_6(\mathbf{t}^{(n)}) \&  \tau>\epsilon_2$ }
 		\STATE Obtain $\mathbf{t}_{\text{temp}}^{(n)} \triangleq \mathbf{t}_m^{(n)} + \tau \nabla_{\mathbf{t}_{m}} f_6(\mathbf{t}_m^{(n)})$.
 		\STATE Obtain $\mathbf{t}_{\text{temp}}^{n+1}$ via (\ref{p9}). 
 		\STATE Update step size $\tau=\frac{1}{2}\tau$.
 		\ENDWHILE	
 		\ENDFOR
 		\STATE Calculate the sum rate $R^{n+1}$ using (\ref{sum}).
 		\STATE Update $n= n+1$.
 		\ENDWHILE
 		\STATE \textbf{Return:} $\mathbf{W}$, $\mathbf{V}$ and $\mathbf{t}$.
 	\end{algorithmic}
 	
 \end{algorithm}
  Specifically, the $m$-th reference point update rule for the $n$-th iteration is as follows:
 \begin{equation}\label{p9}
 	\mathbf{t}_{m}^{(n+1)}\triangleq\left\{\begin{array}{l}
 		\mathbf{t}_{\text{temp}}^{(n)}, \text { if } \mathbf{t}_{\text{temp}}^{(n)} \in \mathcal{C}_{m} \text { and } f_6(\mathbf{t}_{\text{temp}}^{(n)}) \geq f_6(\mathbf{t}^{(n)}), \\
 		P_{\mathcal{C}_m}(\mathbf{t}_{\text{temp}}^{(n)}), \text { otherwise } ,
 	\end{array}\right.
 \end{equation}
 where $\mathbf{t}_{\text{temp}}^{(n)} \triangleq \mathbf{t}_m^{(n)} + \tau \nabla_{\mathbf{t}_{m}} f_6(\mathbf{t}_m^{(n)})$, with $\tau$ representing the step size. An important property of gradient ascent's update rule is its careful choice of step sizes $\tau$. Moreover, the projection method ensures that each iteration of optimization process moves towards feasible solutions, thereby enhancing convergence efficiency.

The proposed AO algorithm is summarized in Algorithm \ref{alg1},
 which serves as a benchmark for evaluating the performance of MA-assisted near-field communication systems utilizing a subarray-connected structure.
 Subsequently, a brief analysis of the convergence of the proposed algorithm is presented. 
 Specifically, for the $(n+1)$-th iteration, when $\boldsymbol{\eta}^{n+1}$ and $\boldsymbol{\mu}^{n+1}$ are fixed, $f_1(\mathbf{W}^{n+1}, \mathbf{V}^{n+1}, \mathbf{t}^{n+1}, \boldsymbol{\eta}^{n+1}, \boldsymbol{\mu}^{n+1})=f_1(\mathbf{W}^{n+1}, \mathbf{V}^{n+1}, \mathbf{t}^{n+1})$, we have
\begin{equation}
	\begin{aligned}
		& f_1(\mathbf{W}^{n+1}, \mathbf{V}^{n+1}, \mathbf{t}^{n+1}, \boldsymbol{\eta}^{n+1}, \boldsymbol{\mu}^{n+1}) \\
		& \geq f_5(\mathbf{W}^{n+1}, \mathbf{V}^{n+1}, \mathbf{t}^{n+1}, \boldsymbol{\eta}^{n}, \boldsymbol{\mu}^{n}) \\
		& \geq f_3(\mathbf{W}^{n+1}, \mathbf{V}^{n+1}, \mathbf{t}^{n}, \boldsymbol{\eta}^{n}, \boldsymbol{\mu}^{n}) \\
		& \geq f_2(\mathbf{W}^{n}, \mathbf{V}^{n+1}, \mathbf{t}^{n}, \boldsymbol{\eta}^{n}, \boldsymbol{\mu}^{n}) \\
		& \geq f_1(\mathbf{W}^{n}, \mathbf{V}^{n}, \mathbf{t}^{n}, \boldsymbol{\eta}^{n}, \boldsymbol{\mu}^{n}),
	\end{aligned}
\end{equation}
by similar reasoning, given fixed $\boldsymbol{\eta}^{n}$ and $\boldsymbol{\mu}^{n}$, our analysis demonstrates that the objective function $f_1$ is non-decreasing, satisfying $f_1(\mathbf{W}^{n+1}, \mathbf{V}^{n+1}, \mathbf{t}^{n+1}) \geq f_1(\mathbf{W}^{n}, \mathbf{V}^{n}, \mathbf{t}^{n})$. This convergence is theoretically ensured by the system's power constraints and will be validated by the subsequent simulation results.

 \section{Simulation  Results}\label{S4}

 \begin{figure}[t]
	\begin{center}
	\includegraphics[scale=0.545]{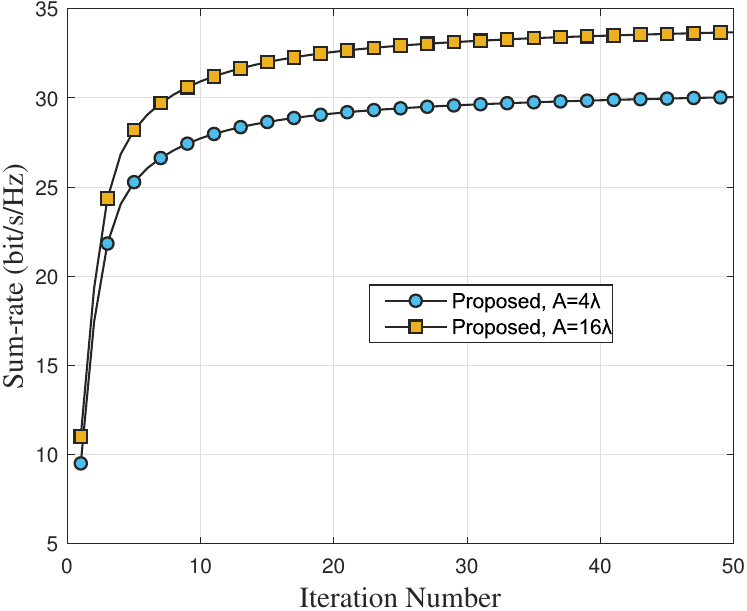} 
	\caption{The sum-rate versus the number of iterations.} 
	\label{fig:a}
	\end{center}
\end{figure}

\begin{figure}[t]
	\begin{center}
		\includegraphics[scale=0.545]{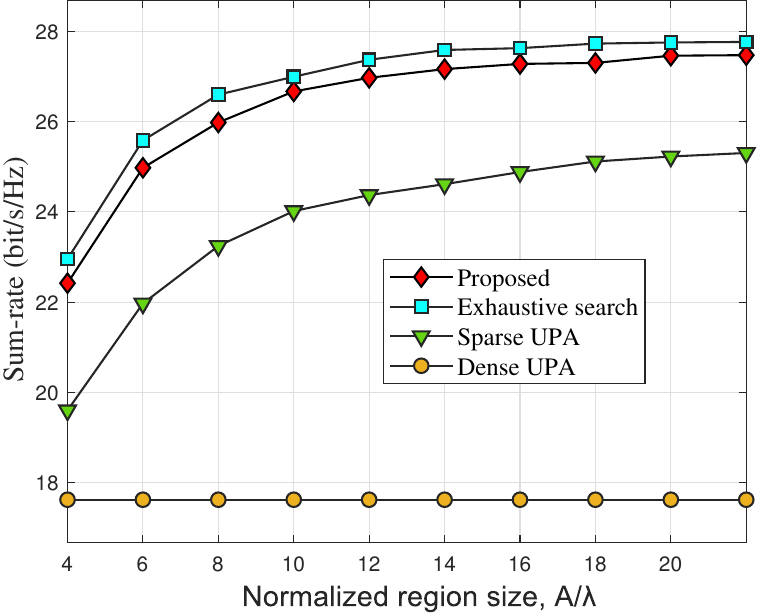} 
			\caption{Sum-rate versus the normalized region size} 
		\label{fig:b}
	\end{center}
\end{figure}

\begin{figure}[t]
	\begin{center}
		\includegraphics[scale=0.545]{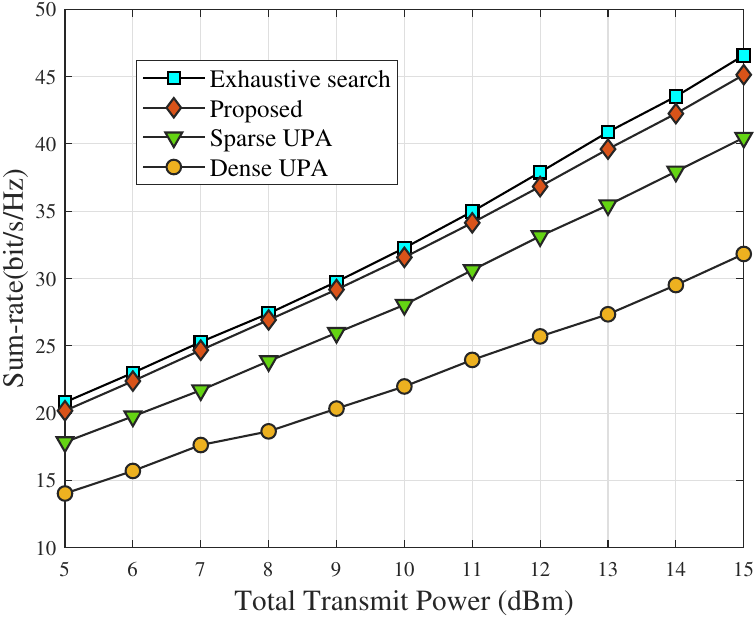} 
		\caption{Sum-rate versus transmit power.} 
		\label{fig:c}
	\end{center}
\end{figure}

 	
 	
 
 In this section, we conduct numerical simulations to validate the effectiveness of our proposed algorithms. We investigate a setup where a 64-antenna BS at $f_c$ = 30 GHz provides service to $K$ = 16 users. Unless specified otherwise, each subarray consists of $N$ = 4
 antennas with inter-antenna spacing $\frac{\lambda}{2} $. The 2D MA array is a square of dimensions $A \times A$, where $A = 20\lambda$. For each MSA, the movable frame has dimensions of $\frac{A}{4} \times \frac{A}{4}$.
The distance between the BS and the $k$-th user is uniformly distributed from 5 to 30 m. In addition, $P$ = 10 dBm, $N_p$ = 6, $\sigma^2_k$ = $-80$ dBm, $\rho = 30$, $\tau^0$ = 0.1, $\epsilon_1$ = $\epsilon_2$ = $\epsilon_3$ = $10^{-3}$, $I$ = 200. All the results 
 are averaged over $10^3$ Monte Carlo simulations. The proposed algorithm is evaluated against the following benchmark schemes: (1) \textbf{Exhaustive search}: the MSA moving region is discretized into multiple grid points with a step size of $\frac{\lambda}{100} \times \frac{\lambda}{100}$, and the optimal MSA position is determined by searching these points. (2) \textbf{Sparse UPA}: A large $4 \times 4$ MA array is constructed from 16 sub-arrays, each a $2 \times 2$ UPA, uniformly arranged with an inter-sub-array spacing of $\frac{A}{8}$. (3) \textbf{Dense UPA}: An $8 \times 8$ UPA is deployed at the BS, with an inter-antenna spacing of $\frac{\lambda}{2}$. For both the \textbf{Sparse} and \textbf{Dense UPA} baselines, the hybrid beamforming matrices are designed by directly extending the far-field AO approach proposed in \cite{10161727} to the considered near-field channel model.

 Fig. \ref{fig:a} illustrates the convergence behavior of the proposed AO algorithm under different MA region sizes. It is observed that the sum-rate increases monotonically and achieves steady-state convergence within approximately 30 iterations. Notably, the convergence rate remains robust to variations in the normalized region size $A/\lambda$. This stability demonstrates that the proposed framework is efficient and reliable, thereby validating the theoretical convergence analysis in Section \ref{S3}.

  Fig. \ref{fig:b} depicts the sum-rate versus the normalized region size $A/\lambda$ for various schemes. It is evident that the proposed scheme consistently outperforms all FPA-based benchmarks, with the sole exception of the exhaustive search. Specifically, compared to the Sparse UPA scheme, the proposed MSA-aided architecture significantly improves the sum-rate by more effectively leveraging the additional spatial DoFs provided by the MA's mobility. The sum-rate is observed to increase with the expansion of the region size before ultimately reaching a stable state. This phenomenon is primarily attributed to the enhanced spatial resolution afforded by MAs; by proactively optimizing antenna positions, the system effectively decorrelates multiuser channels and drives them toward near-orthogonality, thereby suppressing multiuser interference to near its theoretical lower bound. Moreover, as the region size increases, the gradient-based method may converge to a local optimum, leading to a noticeable performance gap between the proposed scheme and the exhaustive search. Nevertheless, by avoiding the prohibitive complexity of exhaustive grid searching, the proposed scheme strikes an excellent trade-off between computational efficiency and sum-rate performance.
  
 Fig. \ref{fig:c} shows the relationship between sum-rate and transmit power. As expected, the sum-rate of all schemes scales with $P$, yet the proposed MSA-aided HBF architecture maintains a superior growth slope compared to both Sparse UPA and Dense UPA. This superiority stems from the synergistic effect of the enlarged array aperture and the position-dependent DoFs. Specifically, unlike conventional FPA systems that are limited by fixed spatial sampling, the MA-enabled system can dynamically reshape the effective NF steering vectors, maximizing the beamfocusing gain at the user locations while minimizing signal leakage into interfering directions.

 


\section{Conclusions}\label{Con} 

 In this paper, we investigated an MSA-aided NF MU-MISO communication system and developed an AO algorithm to jointly optimize digital beamforming, analog beamforming, and MSA positions for sum-rate maximization. Numerical results confirm that the proposed MSA-aided NF architecture synergistically harnesses both position-dependent and near-field distance-dependent DoFs, leading to significant enhancements in spatial resolution and spectral efficiency. Compared to existing MA-enabled near-field designs that often neglect HBF or simplify array architectures, our framework offers a more practical and high-performance solution for XL-MIMO scenarios, substantially outperforming conventional FPA systems. Future research could extend this MSA-aided HBF framework to integrated sensing and communication (ISAC) scenarios, further exploiting the high-resolution beamfocusing for precise target localization.
  
\begin{appendices} 

\end{appendices}

\bibliographystyle{IEEEtran}
\bibliography{reference.bib}

\vspace{12pt}

\end{document}